\documentclass[conference]{IEEEtran}
\IEEEoverridecommandlockouts
\usepackage[letterpaper, left=0.760in, right=0.676in, top=1in, bottom=1.4in]{geometry}

\usepackage{cite}
\usepackage{amsmath,amssymb}
\usepackage{algorithmic}
\usepackage{graphicx}
\usepackage{textcomp}
\usepackage{xcolor}
\usepackage{url}
\usepackage{bm}
\usepackage{booktabs}
\usepackage{algorithm}
\usepackage{algorithmic}
\def\BibTeX{{\rm B\kern-.05em{\sc i\kern-.025em b}\kern-.08em
		T\kern-.1667em\lower.7ex\hbox{E}\kern-.125emX}}

\begin{document}
	%
	%
	\title{Simultaneous Wireless Information and Power Transfer for Fluid Antenna Systems}
	
	\author{
		\IEEEauthorblockN{
			Feilong Zhang\IEEEauthorrefmark{1}, \quad
			Jianxin Dai\IEEEauthorrefmark{1}\IEEEauthorrefmark{2}, \quad
			Zhaohui Yang\IEEEauthorrefmark{3}\IEEEauthorrefmark{4}, \quad
			Kai-Kit Wong\IEEEauthorrefmark{5}\IEEEauthorrefmark{6}, \quad
			Lingyuxiu Li\IEEEauthorrefmark{1}, \quad
			and Jianglin Ye\IEEEauthorrefmark{1}
		}
		\IEEEauthorblockA{
			$\IEEEauthorrefmark{1}$College of Science, Nanjing University of Posts and Telecommunications, Nanjing, China\\
			$\IEEEauthorrefmark{2}$National Mobile Communications Research Laboratory, Southeast University, Nanjing, China\\
			$\IEEEauthorrefmark{3}$College of Information Science and Electronic Engineering, Zhejiang University, Hangzhou, China\\
			$\IEEEauthorrefmark{4}$Zhejiang Provincial Key Laboratory of Info. Proc., Commun. \& Netw. (IPCAN), Hangzhou, China\\
			$\IEEEauthorrefmark{5}$Department of Electronic and Electrical Engineering, University College London, Torrington Place, United Kingdom\\
			$\IEEEauthorrefmark{6}$Yonsei Frontier Laboratory and School of Integrated Technology, Yonsei University, Seoul, Korea\\[0.5em]  
			E-mails: 1022082113@njupt.edu.cn, daijx@njupt.edu.cn, yang\_zhaohui@zju.edu.cn, \\
			\hspace{2.2em}kai-kit.wong@ucl.ac.uk, 1022082112@njupt.edu.cn, 1222087520@njupt.edu.cn
		}
	}
	\maketitle

\begin{abstract}
Fluid antenna is a promising wireless communication technology that enhances communication rate by changing the antenna positions. This article proposes a new communication system that combines multiple-input single-output (MISO) fluid antennas with traditional fixed-position antennas, utilizing antenna position optimization to improve energy harvesting efficiency. In this model, we consider simultaneous wireless information and power transfer (SWIPT) which transmits identical signals from the base station to both information receiver (IR) and energy receiver (ER). We strive to enhance the power delivered to the ER by fine-tuning the positions of transmit and receive fluid antennas, along with optimizing the transmit covariance matrix, subject to a given minimum signal-to-interference-plus-noise ratio (SINR) constraint at the IR. Simulation results indicate that fluid antenna systems significantly enhance the energy harvesting efficiency of the ER compared to traditional fixed-position antennas.
\end{abstract}
\IEEEpeerreviewmaketitle
\section{Introduction}
\IEEEPARstart{I}{n} the past few decades,  wireless communication has developed rapidly and has significantly changed our lives. Throughout the development of wireless communication systems, enhancing communication rate has always been a primary goal. To improve communication efficiency, fluid antenna systems (FASs) capable of flexibly adjusting antenna positions along one-dimensional (1D) lines have been proposed  \cite{b1}, \cite{b2}. Using conductive fluid as the antenna material makes it possible to freely move the position of the receive antennas among all potential access points distributed along a fixed, finite-length line, thereby communication rate is enhanced. However, the restrictions imposed by liquid materials dictate that FAS can solely accommodate a singular fluid antenna, which is capable of movement solely along a linear path, its capacity to fully capitalize on spatial variations in the wireless channel is constrained. In order to fully propose more spatial degrees of freedom (DoFs) to enhance communication rate, the movable antennas (MAs) were proposed in \cite{b3}. Using flexible cables to connect MAs to the radio frequency (RF) chains  allows real time adjustment of antenna position by mechanical driving \cite{b4},\cite{b5},\cite{b6}. Unlike fixed-position antenna systems, this system reshapes the channel matrix between antennas by changing the position of the receive/transmit antennas, thus achieving a higher communication rate\cite{b7},\cite{b8}.\\
\indent Simultaneous wireless information and power transfer (SWIPT) is an integration of combining wireless information transfer (WIT) and wireless power transfer (WPT). SWIPT leverages the inherent properties of RF signals, enabling concurrent transmission of both information and energy \cite{b9},\cite{b10}.
In 2008, L. R. Varshney introduced the concept of the energy-information rate trade-off in binary and Gaussian channels, approaching it through the lens of information theory in \cite{b11},\cite{b12},\cite{b13}, proposing the concept of SWIPT. Currently, SWIPT technology falls into two categories: the first where the receiving end simultaneously performs information decoding and energy harvesting (EH), and the transmitter carries both information and energy in the RF signal for transmission \cite{b14},\cite{b15},\cite{b16}, the second category involves separate functions at the receiving end for information decoding and EH. This implies that there are recipients for information and beneficiaries for energy collection, with the transmitter transmitting separate signals for information and energy\cite{b17},\cite{b18},\cite{b19}. We assume that perfect channel state information(CSI) is available at both the transmitter and receiver. \\
\indent The rest of the article is organized as follows: Section II introduces the system model and formulates the energy harvesting problem. Section III provides an alternating optimization algorithm to solve the formulated problem. Section IV presents numerical results and discussion. Finally, Section V provides a summary of this article.\\

\begin{figure}[h]
\centering
\includegraphics[width=6cm,height=6cm]{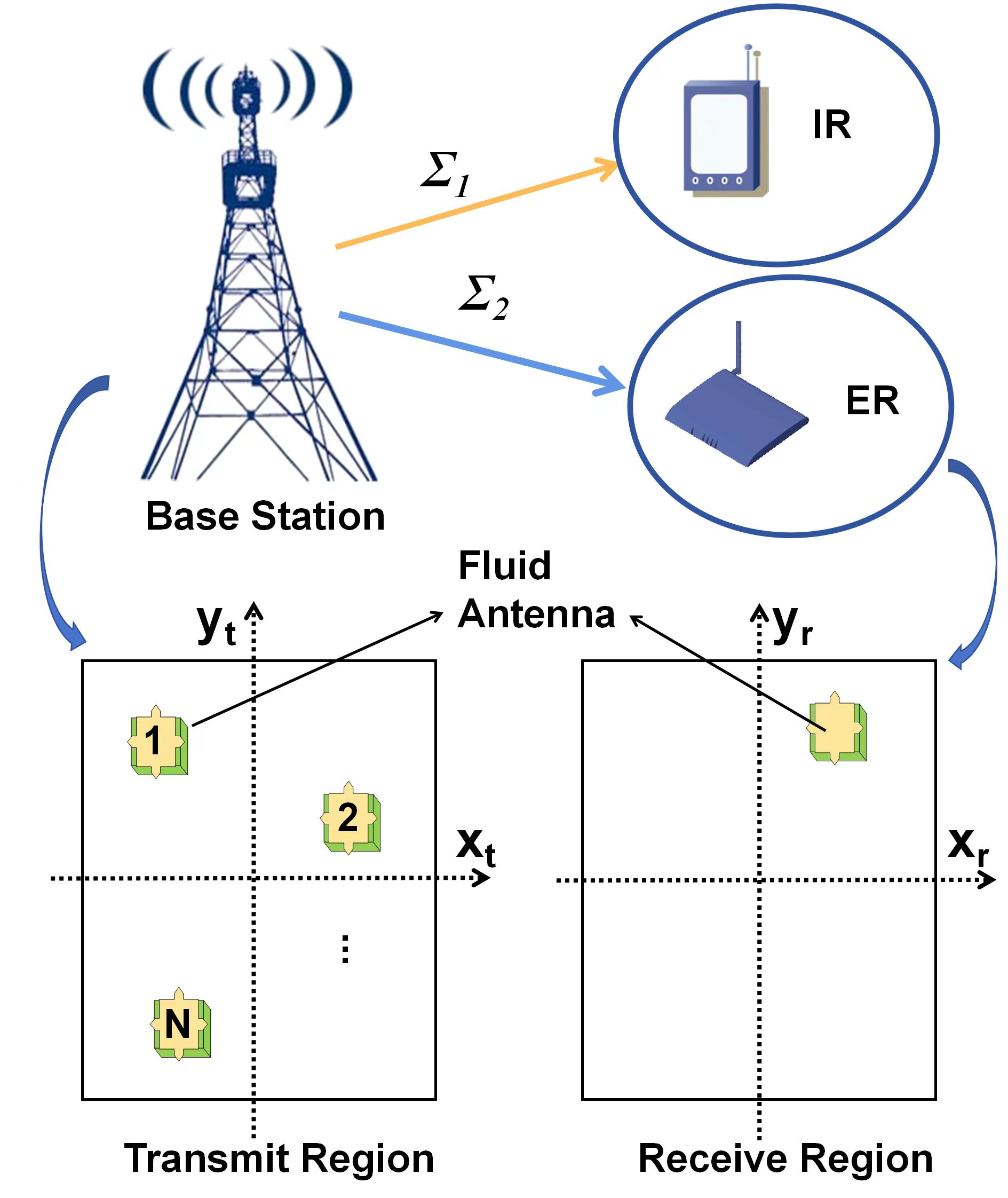}
\caption {A MISO fluid antenna system for dual information and energy transfer.}
\label {1}
\end{figure}
\section{System Model}
\subsection{Fluid Antenna MISO System}
 As shown in Fig.~\ref{1}, this paper considers a three-node wireless multiple-input single-output (MISO) system, base station equipped with $N$ fluid antennas (FAs), a single FA at the ER for EH, and an IR  with a single traditional fixed-position antenna. The FAs on the base station and ER are interfaced with the RF chains via cables, permitting real-time adjustment of their positions. Define the Cartesian coordinate $\bm{t}_{n}=[x_{t,n},y_{t,n}]^{T}\in C_{t}$ as the $n$th $(n=1,2,... N)$ transmit FA position, $\bm{r}=[x_{r}, y_{r}]^{T}\in C_{r}$ denotes receive FA position. The coordinates of the IR can be set to $\bm{r}_{0}=[x_{0},y_{0}]^{T}$. $C_{t}$ represents the two-dimensional (2D) region of the transmission end, and $C_{r}$  represents the two-dimensional (2D) region of the reception end, where FAs can move freely within these regions.

The paper considers narrow-band channels with slow fading and focuses on a quasi-static fading block. For the MISO communication
system with FAs, it can obtain a new channel configuration when the positions of the receive and transmit FAs are adjusted.
 Let $\bm{t}=[t_{1},t_{2},\ldots,t_{N}]\in R^{2\times N}$ represent the coordinate set of $N$ transmit FAs. Then, the MISO channel vector from the transmitter to the receiver is given by
$\bm{h}(\bm{t},\bm{r})\in C^{1\times N}$, which is a function of $\bm{t}$ and $\bm{r}$ in general. The transmit signal vector is defined as $\bm{s}\in C^{N}$ , and the covariance matrix of the transmit signal is denoted by $\bm{Q}\triangleq \mathrm{E}\{\bm{s}\bm{s}^{H}\}\in \textbf{C}^{N\times N}$, $\bm{Q}\succeq 0$. We assume an average power constraint at the base station, represented as Tr$(\bm{Q})\leq P$. Therefore, the signal from the
transmitter to the receiver can be modeled by
\begin{equation}
y(\bm{t},\bm{r})=\bm{h}(\bm{t},\bm{r})\bm{s}+z,
\end{equation}
where $z\sim \mathcal{CN}(0,\sigma^{2})$ is an additive white Gaussian noise (AWGN) vector with power $\sigma^{2}$.

\subsection{Field-Response Based Channel Model}
For MISO communication systems with FAs, the channel vectors depend on both the environmental conditions of signal transmission and FAs' positions. In this context, we assume a long-distance transmission model, where the dimensions of base station and ER are insignificant compared to the distance over which the signal propagates. Therefore, in the transmit/receive region, each channel path component exhibits identical angle of departure (AoD) or angle of arrival (AoA) and complex path coefficients, the phases of the complex path coefficients differ across various transmit/receive antennas pairs positioned at different locations.

In \cite{b3}, denote the number of transmit paths and receive paths as $L_{t}$ and $L_{r}$, the field-response vector of one transmit FA is
\begin{equation}
\bm{g}(\bm{t})\triangleq[e^{j\frac{2\pi}{\lambda}\rho_{t}^{1}(\bm{t})},e^{j\frac{2\pi}{\lambda}\rho_{t}^{2}(\bm{t})},\cdots,e^{j\frac{2\pi}{\lambda}\rho_{t}^{L_{t}}(\bm{t})}]^{T}\in C^{L_{t}}.
\end{equation}

The field-response matrix for the base station region is
\begin{equation}
\bm{G}(\bm{t})\triangleq[\bm{g}(\bm{t}_{1}),\bm{g}(\bm{t}_{2}),\cdots,\bm{g}(\bm{t}_{N})]\in C^{L_{t}\times N}.
\end{equation}
\indent Similarly, the field-response vector for the receive region is
\begin{equation}
\bm{f}(\bm{r})\triangleq[e^{j\frac{2\pi}{\lambda}\rho_{r}^{1}(\bm{r})},e^{j\frac{2\pi}{\lambda}\rho_{r}^{2}(\bm{r})},\cdots,e^{j\frac{2\pi}{\lambda}\rho_{r}^{L_{r}}(\bm{r})}]^{T}\in C^{L_{r}}.
\end{equation}
We establish a path response matrix $\mathbf{\Sigma}\in C^{L_{r}\times L_{t}}$ to denote the path response from the transmit regions origin to that of the receive region. The path response matrix for the $p$th transmit path and $q$th receive path is denoted by $\mathbf{\Sigma} [q,p]$.
The channel from the base station to the receiver is
\begin{equation}
\bm{h}(\bm{t},\bm{r})^{H}=\bm{f}(\bm{r})^{H}\mathbf{\Sigma} \bm{G}(\bm{t}).
\end{equation}
\indent The received signal at the IR is
\begin{equation}
y_{I}=\bm{h}_{I}(\bm{t},\bm{r}_{0})\bm{s}+z_{I},
\end{equation}
where $z_{I}\sim \mathcal{CN}(0,\sigma_{I}^{2})$ stands for the additive white Gaussian noise at IR with power $\sigma_{I}^{2}$.

The IR signal-to-interference-plus-noise ratio (SINR) can be expressed as
\begin{equation}
\gamma=\frac{\lvert \bm{h}_{I}(\bm{t},\bm{r}_{0})\bm{s}\bm{s}^{H}\bm{h}_{I}(\bm{t},\bm{r}_{0})^{H} \rvert^{2}}{\sigma_{I}^{2}}.
\end{equation}

 The received signal at the ER is
\begin{equation}
y_{E}=\bm{h}_{E}(\bm{t},\bm{r})\bm{s}+z_{E},
\end{equation}
where $z_{E}\sim \mathcal{CN}(0,\sigma_{E}^{2})$ stands for the additive white Gaussian noise at ER with power $\sigma_{E}^{2}$.

The power harvested by the EH is given by
\begin{equation}
\mathrm{W}=\eta \mathrm{tr}(\bm{h}_{E}(\bm{t},\bm{r})\bm{Q} \bm{h}_{E}(\bm{t},\bm{r})^{H}),
\end{equation}
where $\eta$ represents the efficiency of EH. For the purposes of this paper, we assume a value of 1 for $\eta$.

\subsection{Problem Formulation}
To prevent interference among transmit antennas, it is required that the minimum distance between any two antennas at the transmit region should be set to $D$, i.e., $\lVert \bm{t}_{k}-\bm{t}_{l} \rVert_{2} , k,l=1,2,... N,k\neq l$ \cite{b3}. Then, our objective is to maximize the EH through joint optimization of the transmit covariance $\bm{Q}$, transmit FA position $\bm{t}$, and receive position $\bm{r}$, subject to constraints on the minimizing SINR at the IR, minimum distance constraints on FA positions, and transmitter power constraints. For optimizing the problem, we can formulate as
\begin{subequations}\label{P1}
\begin{align}
(\mathrm{P1})~~~  \max_{\bm{t},\bm{r},\bm{Q}}&~\mathrm{tr}(\bm{h}_{E}(\bm{t},\bm{r})\bm{Q} \bm{h}_{E}(\bm{t},\bm{r})^{H})
\tag{\theequation}\\
\text{s.t.}
&\quad \bm{t}\in C_{t},\label{P1:sub1}\\
&\quad \bm{r}\in C_{r},\label{P1:sub2}\\
&\quad \lVert \bm{t}_{k}-\bm{t}_{l} \rVert_{2}\geq D,~k,l=1,2,\ldots,N,~k\neq l,\label{P1:sub3}\\
&\quad \frac{\lvert \bm{h}_{I}(\bm{t},\bm{r}_{0})\bm{s} \rvert^{2}}{\sigma_{I}^{2}} \geq \bar{\gamma},\label{P1:sub4}\\
&\quad \operatorname{tr}(\bm{Q})\leq \mathrm{P},\label{P1:sub5}\\
&\quad \bm{Q}\succeq 0.\label{P1:sub6}
\end{align}
\end{subequations}

Problem $(\mathrm{P1})$ is evidently non-convex because it is non-convex at the FA positions $\bm{t}$ and $\bm{r}$, and both \eqref{P1:sub3} and \eqref{P1:sub4} are non-convex. Furthermore, $\bm{Q}$ is associated with both $\bm{t}$ and $\bm{r}$, which makes the problem difficult to solve.
\section{Proposed Algorithm}
We present an alternating optimization algorithm designed to address the Problem $(\mathrm{P1})$. Firstly, the optimization variables of the objective function are $\{\bm{t}_{n}\}_{n=1}^{N}\cup\{\bm{r}\}\cup\bm{Q}$, which facilitates our optimization. Then, to solve each of the three sub-problems, we optimize $\bm{Q}$, the energy receiver FA position $\bm{r}$, and one base station FA position $\bm{t}_{n}$, keeping the other variables constant. It can solve these three sub-problems, yielding at least one locally optimal solution that satisfies the conditions.
\subsubsection {Optimization of $\bm{Q}$ } To optimize $\bm{Q}$ , we give the transmit FA positions $\{\bm{t}_{n}\}_{n=1}^{N}$ and the receive FA position ${\bm{r}}$. The optimization problem can be represented as
\begin{subequations}\label{P2}
\begin{align}
\max_{\bm{Q}}&~\mathrm{tr}(\bm{h}_{E}(\bm{t},\bm{r})\bm{Q} \bm{h}_{E}(\bm{t},\bm{r})^{H})
\tag{\theequation}\\
\text{s.t.}
&\quad \frac{\mathrm{tr}(\bm{h}_{I}(\bm{t},\bm{r}_{0})\bm{Q} \bm{h}_{I}(\bm{t},\bm{r}_{0})^{H})}{\sigma_{I}^{2}}-\bar{\gamma}\geq 0,\label{P2:sub2}\\
&\quad \operatorname{tr}(\bm{Q})\leq \mathrm{P},\label{P2:sub2}\\
&\quad \bm{Q}\succeq 0.\label{P2:sub3}
\end{align}
\end{subequations}
\indent This is a convex optimization problem about $\bm{Q}$, and the optimal value can be effectively solved through the CVX toolbox \cite{b20}.

\subsubsection {Optimization of  ${\bm{r}}$} To optimize the receive FA position ${\bm{r}}$, we fix the transmit FA positions $\{\bm{t}_{n}\}_{n=1}^{N}$ and $\bm{Q$}. Specifically, we establish a matrix that is positive definite
\begin{subequations}\label{A}
\begin{align}
\bm{A}_{n}=\bm{\Sigma}_{E}\bm{G}(\bm{t})\bm{Q}\bm{G}(\bm{t})^{H}\bm{\Sigma}_{E}^{H}\tag{\theequation}.
\end{align}
\end{subequations}
 Notice that $\bm{A}_{n}$ is a constant matrix regardless of $\bm{r}$. Consequently, we can obtain the following
\begin{subequations}\label{P0}
\begin{align}
(\mathrm{P2})~~~\max_{\bm{r}}&~\bm{f}(\bm{r})^{H}\bm{A}_{n}\bm{f}(\bm{r})
\tag{\theequation}\\
\text{s.t.}
&\quad \bm{r}\in C_{r}.\label{P0:sub1}
\end{align}
\end{subequations}
One can see that the objective function exhibits non-concave with respect to $\bm{r}$. To solve the above problem, we utilize the successive convex approximation (SCA) method, which can optimize the position of the receive antenna $\bm{r}$ . The objective function is convex with respect to $\bm{f}(\bm{r})$. A convex function can always be approximated from below by its first-order Taylor expansion at any point. In the $i$th iteration of SCA, we can obtain a lower bound for $\bm{f}(\bm{r})^{H}\bm{A}_{n}\bm{f}(\bm{r})$ as
\begin{subequations}
\begin{align}
x(\bm{r})&= \bm{f}(\bm{r})^{H}\bm{A}_{n}\bm{f}(\bm{r})\tag{\theequation}\\
&\geq \bm{f}(\bm{r}^{i})^{H}\bm{A}_{n}\bm{f}(\bm{r}^{i})+2\mathrm{Re}\{\bm{f}(\bm{r}^{i})^{H}\bm{A}_{n}(\bm{f}(\bm{r})-\bm{f}(\bm{r}^{i}))\}\notag\\
&= 2\mathrm{Re}\{\bm{f}(\bm{r}^{i})^{H}\bm{A}_{n}\bm{f}(\bm{r})\}-\bm{f}(\bm{r}^{i})^{H}\bm{A}_{n}\bm{f}(\bm{r}^{i}),\notag
\end{align}
\end{subequations}
where $\bm{r}^{i}$ is a constant vector representing the value of $\bm{r}$ in the $i$th iteration, and $\bm{f}(\bm{r}^{i})^{H}\bm{A}_{n}\bm{f}(\bm{r})$ is a constant term. Consequently, maximizing $\bm{f}(\bm{r})^{H}\bm{A}_{n}\bm{f}(\bm{r})$ is equivalent to maximizing $\bar{x}(\bm{r})=\mathrm{Re}\{\bm{f}(\bm{r}^{i})^{H}\bm{A}_{n}\bm{f}(\bm{r})\}$. However, it is not a convex function with respect to $\bm{r}$. Therefore, it is insufficient to construct a lower bound proxy function for the objective function only based on the first-order Taylor expansion of $\bar{x}(\bm{r})$. We construct a local approximation of the objective function using the second-order Taylor expansion. Then we denote the gradient vector of and the Hessian matrix of $\bar{x}(\bm{r})$ over $\bm{r}$ by $\nabla\bar{x}(\bm{r})\in R^{2}$ and $\nabla^{2}\bar{x}(\bm{r})\in R^{2\times2}$, and construct positive real number $\delta$, such that $\delta \bm{I}_{2}\succeq \nabla^{2}\bar{x}(\bm{r})$. Please refer expansion to Appendix B in \cite{b3} for the proof. Hence, according to the Taylor theorem, we can obtain the following bound is $\bar{x}(\bm{r})$
\begin{subequations}
\begin{align}
\bar{x}(\bm{r})=& \mathrm{Re}\{\bm{f}(\bm{r}^{i})^{H}\bm{A}_{n}\bm{f}(\bm{r})\}\tag{\theequation}\\
\geq & \bar{x}(\bm{r}^{i})+\nabla\bar{x}(\bm{r}^{i})^{T}(\bm{r}-\bm{r}^{i})-\frac{\delta}{2}(\bm{r}-\bm{r}^{i})^{T}(\bm{r}-\bm{r}^{i})\notag\\
=& -\frac{\delta}{2}\bm{r}^{T}\bm{r}+(\nabla\bar{x}(\bm{r}^{i})+\delta \bm{r}^{i})^{T}\bm{r}-\nabla\bar{x}(\bm{r}^{i})^{T}\bm{r}^{i}\notag\\
&+\bar{x}(\bm{r}^{i})-\frac{\delta}{2}(\bm{r}^{i})^{T}\bm{r}^{i}\notag,
\end{align}
\end{subequations}
where $-\nabla\bar{x}(\bm{r}^{i})^{T}\bm{r}^{i}+\bar{x}(\bm{r}^{i})-\frac{\delta}{2}(\bm{r}^{i})^{T}\bm{r}^{i}$ is a constant, maximizing $\bar{x}(\bm{r})$ is equivalent to maximizing $\tilde{x}(\bm{r})\triangleq-\frac{\delta}{2}\bm{r}^{T}\bm{r}+(\nabla\bar{x}(\bm{r}^{i})+\delta \bm{r}^{i})^{T}\bm{r}$. We can obtain the following
\begin{subequations}\label{P3}
\begin{align}
(\mathrm{P3})~~~\max_{\bm{r}}&~-\frac{\delta}{2}\bm{r}^{T}\bm{r}+(\nabla\bar{x}(\bm{r}^{i})+\delta \bm{r}^{i})^{T}\bm{r}
\tag{\theequation}\\
\text{s.t.}
&\quad \bm{r}\in C_{r}.\label{P3:sub1}
\end{align}
\end{subequations}

\indent The objective function is a concave quadratic function on $\bm{r}$, and without considering constraints \eqref{P3:sub1}, the maximum global optimal solution is
\begin{equation}
\begin{split}
\bm{r}^{\ast}_{i+1}=\frac{1}{\delta}\nabla\bar{x}(\bm{r}^{i})+\bm{r}^{i}
\end{split}
\end{equation}
\indent If $\bm{r}^{\ast}_{i+1}$ satisfies \eqref{P3:sub1}, it is the global optimum for problem (P3). If it does not satisfy the constraint conditions, we know that it is a quadratic programming (QP) problem and can obtain the optimal solution through the quadratic programming function.

\addtolength{\topmargin}{0.257in}

\subsubsection {Optimization of $\bm{t}_{n}$} To optimize the transmit $\bm{t}_{n}$, we give $\bm{Q}$, $\{\bm{r}\}$, and $\{\bm{t}_{l},l\neq n\}_{l=1}^{N}$ , $\forall_{n}\in \mathcal{N}=\{1,2,\cdots N\}$. For the objective function Problem $(\mathrm{P1})$, we can transform it into
\begin{subequations}
\begin{align}
&\mathrm{tr}(\bm{h}_{E}(\bm{t},\bm{r})\bm{Q}\bm{h}_{E}(\bm{t},\bm{r})^{H})\tag{\theequation}\\
=&\mathrm{tr}(\bm{h}_{E}(\bm{t},\bm{r})\bm{s}\bm{s}^{H}\bm{h}_{E}(\bm{t},\bm{r})^{H})=\mathrm{P}\mathrm{tr}(\bm{h}_{E}(\bm{t},\bm{r})\bm{h}_{E}(\bm{t},\bm{r})^{H}).\notag
\end{align}
\end{subequations}
The problem can be equivalent to maximizing
\begin{equation}
\begin{split}
\mathrm{tr}(\bm{h}_{E}(\bm{t},\bm{r})\bm{h}_{E}(\bm{t},\bm{r})^{H}).
\end{split}
\end{equation}
Based on this, we define the $n$th element of $\bm{h}_{E}(\bm{t},\bm{r})\in C^{1\times N}$ as $\bm{h}_{E,n}(\bm{t}_{n})=\bm{f}(\bm{r})^{H}\mathbf{\Sigma}_{E}\bm{g}(\bm{t}_{n})$, then, we remove $\bm{h}_{E,n}(\bm{t}_{n})$ from $\bm{h}_{E}(\bm{t},\bm{r})\in C^{1\times N}$ and denote the remaining sub-vector:
\begin{subequations}
\begin{align}
&\bm{h}_{E}^{\bm{t}}=[\bm{h}_{E,1}(\bm{t}_{1}),\bm{h}_{E,2}(\bm{t}_{2}),\cdots , \bm{h}_{E,n-1}(\bm{t}_{n-1}),\bm{h}_{E,n+1}(\bm{t}_{n+1})\notag\\
&,\cdots,\bm{h}_{E}(\bm{t}_{N})]\in C^{1\times(N-1)}.\tag{\theequation}
\end{align}
\end{subequations}
Thus, we can obtain the following
\begin{subequations}
\begin{align}
&\mathrm{tr}(\bm{h}_{E}(\bm{t},\bm{r})h_{E}(\bm{t},\bm{r})^{H})\tag{\theequation}\\
=&\mathrm{tr}(\bm{h}_{E}^{\bm{t}}(\bm{h}_{E}^{\bm{t}})^{H}+\bm{h}_{E,n}(\bm{t}_{n})\bm{h}_{E,n}(\bm{t}_{n})^{H})\notag\\
=&\mathrm{tr}(\bm{h}_{E}^{\bm{t}}(\bm{h}_{E}^{\bm{t}})^{H})+\mathrm{tr}(\bm{h}_{E,n}(\bm{t}_{n})\bm{h}_{E,n}(\bm{t}_{n})^{H}).\notag
\end{align}
\end{subequations}
The problem can be equivalent to maximizing
\begin{subequations}
\begin{align}
\mathrm{tr}(\bm{h}_{E,n}(\bm{t}_{n})\bm{h}_{E,n}(\bm{t}_{n})^{H})=&\mathrm{tr}(\bm{h}_{E,n}(\bm{t}_{n})^{H}\bm{h}_{E,n}(\bm{t}_{n}))\tag{\theequation}\\
=&\bm{g}(\bm{t}_{n})^{H}\mathbf{\Sigma}_{E}^{H}\bm{f}(\bm{r})\bm{f}(\bm{r})^{H}\mathbf{\Sigma}_{E}\bm{g}(\bm{t}_{n}).\notag
\end{align}
\end{subequations}
We establish a matrix that is positive definite
\begin{subequations}\label{B}
\begin{align}
\bm{B}_{n}=\mathbf{\Sigma}_{E}^{H}\bm{f}(\bm{r})\bm{f}(\bm{r})^{H}\mathbf{\Sigma}_{E}.\tag{\theequation}
\end{align}
\end{subequations}
Therefore, we can obtain the following
\begin{subequations}\label{P4}
\begin{align}
(\mathrm{P4})~~~\max_{\bm{t}_{n}}&~\bm{g}(\bm{t}_{n})^{H}\bm{B}_{n}\bm{g}(\bm{t}_{n})
\tag{\theequation}\\
\text{s.t.}
&\quad \bm{t}_{n}\in C_{t},\label{P4:sub1}\\
&\quad \lVert \bm{t}_{n}-\bm{t}_{l} \rVert_{2}\geq D,~l=1,2,\ldots,N,~l\neq n,\label{P4:sub2}\\
&\quad \frac{\mathrm{tr}(\bm{h}_{I}(\bm{t},\bm{r}_{0})\bm{Q} \bm{h}_{I}(\bm{t},\bm{r}_{0})^{H})}{\sigma_{I}^{2}}-\bar{\gamma}\geq 0.\label{P4:sub3}
\end{align}
\end{subequations}
The optimization process is similar to the second sub-problem, we are unable to achieve the optimal solution because of the non-convex
constraints \eqref{P4:sub2} and \eqref{P4:sub3}.

The lower bound proxy function for $\bm{g}(\bm{t}_{n})^{H}\bm{B}_{n}\bm{g}(\bm{t}_{n})$ is
\begin{small}
\begin{subequations}
\begin{align}
y(\bm{t_{n}})&= \bm{g}(\bm{t_{n}})^{H}\bm{B}_{n}\bm{g}(\bm{t_{n}})\tag{\theequation}\\
&\geq\bm{g}(\bm{t}_{n}^{i})^{H}\bm{B}_{n}\bm{g}(\bm{t}_{n}^{i})+2\mathrm{Re}\{\bm{g}(\bm{t}_{n}^{i})^{H}\bm{B}_{n}(\bm{g}(\bm{t_{n}})-\bm{g}(\bm{t}_{n}^{i}))\}\notag\\
&= 2\mathrm{Re}\{\bm{g}(\bm{t}_{n}^{i})^{H}\bm{B}_{n}\bm{g}(\bm{t_{n}})\}-\bm{g}(\bm{t}_{n}^{i})^{H}\bm{B}_{n}\bm{g}(\bm{t}_{n}^{i}).\notag
\end{align}
\end{subequations}
\end{small}
Then, we establish a positive real number denoted as  $\beta_{n}$, such that $\beta_{n} \bm{I}_{2}\succeq \nabla^{2}\bar{y}(\bm{t_{n}})$. Similar to the derivation process in Problem ($\mathrm{P2}$), we can obtain the following
\begin{small}
\begin{subequations}
\begin{align}
\bar{y}(\bm{t}_{n})=& \mathrm{Re}\{\bm{g}(\bm{t}_{n}^{i})^{H}\bm{B}_{n}\bm{g}(\bm{t}_{n})\}\tag{\theequation}\\
\geq & \bar{y}(\bm{t}_{n}^{i})+\nabla\bar{y}(\bm{t}_{n}^{i})^{T}(\bm{t}_{n}-\bm{t}_{n}^{i})-\frac{\beta_{n}}{2}(\bm{t}_{n}-\bm{t}_{n}^{i})^{T}(\bm{t}_{n}-\bm{t}_{n}^{i})\notag\\
=& -\frac{\beta_{n}}{2}\bm{t}_{n}^{T}\bm{t}_{n}+(\nabla\bar{y}(\bm{t}_{n}^{i})+\beta_{n} \bm{t}_{n}^{i})^{T}\bm{t}_{n}-\nabla\bar{y}(\bm{t}_{n}^{i})^{T}\bm{t}_{n}^{i}\notag\\
&+\bar{y}(\bm{t}_{n}^{i})-\frac{\beta_{n}}{2}(\bm{t}_{n}^{i})^{T}\bm{t}_{n}^{i}\notag,
\end{align}
\end{subequations}
\end{small}
 where $-\nabla\bar{y}(\bm{t}_{n}^{i})^{T}\bm{t}_{n}^{i}+\bar{y}(\bm{t}_{n}^{i})-\frac{\beta_{n}}{2}(\bm{t}_{n}^{i})^{T}\bm{t}_{n}^{i}$ is a constant, maximizing $\bar{y}(\bm{t}_{n})$ is equivalent to maximizing $\tilde{y}(\bm{t}_{n})\triangleq-\frac{\beta_{n}}{2}\bm{t}_{n}^{T}\bm{t}_{n}+(\nabla\bar{y}(\bm{t}_{n}^{i})+\beta_{n} \bm{t}_{n}^{i})^{T}\bm{t}_{n}$.

At any given point, the global lower bound of any convex function can be approximated by its first-order Taylor expansion. In the $i$-th iteration of SCA, given the local point margin, we obtain a lower bound of  $\lVert \bm{t}_{n}-\bm{t}_{l} \rVert_{2}\geq D$ as
\begin{equation}
\begin{split}
\frac{1}{\lVert \bm{t}_{n}^{i}-\bm{t}_{l} \rVert_{2}}(\bm{t}_{n}^{i}-\bm{t}_{l})^{T}(\bm{t}_{n}-\bm{t}_{l})\geq D.
\end{split}
\end{equation}
It is a linear constraint about $\bm{t}_{n}$, this proof is the same as the proof in \cite{b3} and is omitted for simplicity.

Next, let us find the lower bound surrogate function for the non-convex constraint (\ref{P4:sub3}):
\begin{subequations}
\begin{align}
&\mathrm{tr}(\bm{h}_{I}(\bm{t},\bm{r}_{0})\bm{Q} \bm{h}_{I}(\bm{t},\bm{r}_{0})^{H})\tag{\theequation}\\
=&\mathrm{tr}(\bm{h}_{I}(\bm{t},\bm{r}_{0})\bm{s}\bm{s}^{H}\bm{h}_{I}(\bm{t},\bm{r}_{0})^{H})=\mathrm{P}\mathrm{tr}(\bm{h}_{I}(\bm{t},\bm{r}_{0})\bm{h}_{I}(\bm{t},\bm{r}_{0})^{H})\notag.
\end{align}
\end{subequations}
The problem can be equivalent to maximizing
\begin{equation}
\begin{split}
\mathrm{tr}(\bm{h}_{I}(\bm{t},\bm{r}_{0})\bm{h}_{I}(\bm{t},\bm{r}_{0})^{H}).
\end{split}
\end{equation}

Based on this, we define the $n$th element of $\bm{h}_{I}(\bm{t},\bm{r}_{0})\in C^{1\times N}$ by $\bm{h}_{I,n}(\bm{t}_{n})=\bm{f}(\bm{r}_{0})^{H}\mathbf{\Sigma}_{I}\bm{g}(\bm{t}_{n})$. Then, we remove $\bm{h}_{I,n}(\bm{t}_{n})$ from $\bm{h}_{I}(\bm{t},\bm{r}_{0})\in C^{1\times N}$ and denote the remaining sub-vector as
\begin{subequations}
\begin{align}
&\bm{h}_{I}^{\bm{t}}=[\bm{h}_{I,1}(\bm{t}_{1}),\bm{h}_{I,2}(\bm{t}_{2}),\cdots , \bm{h}_{I,n-1}(\bm{t}_{n-1}),\notag\\
&\bm{h}_{I,n+1}(\bm{t}_{n+1}),\cdots,\bm{h}_{I}(\bm{t}_{N})]\in C^{1\times(N-1)}.\tag{\theequation}
\end{align}
\end{subequations}

We define a positive definite matrix:
\begin{equation}
\begin{split}
\bm{C}_{n}=\mathbf{\Sigma}_{I}^{H}\bm{f}(\bm{r}_{0})\bm{f}(\bm{r}_{0})^{H}\mathbf{\Sigma}_{I}.
\end{split}
\end{equation}
The derivation process is similar to the objective function of subproblem three, from which we can obtain
\begin{equation}
\begin{split}
\mathrm{tr}(\bm{h}_{I,n}(\bm{t}_{n})\bm{h}_{I,n}(\bm{t}_{n})^{H})=\bm{g}(\bm{t}_{n})^{H}\bm{C}_{n}\bm{g}(\bm{t}_{n}).
\end{split}
\end{equation}
Meanwhile, we can also obtain
\begin{subequations}
\begin{align}
z(\bm{t}_{n})=& \bm{g}(\bm{t}_{n})^{H}\bm{C}_{n}\bm{g}(\bm{t}_{n})\tag{\theequation}\\
\geq & \bm{g}(\bm{t}^{i}_{n})^{H}\bm{C}_{n}\bm{g}(\bm{t}^{i}_{n})+2\mathrm{Re}\{\bm{g}(\bm{t}^{i}_{n})^{H}\bm{C}_{n}(\bm{g}(\bm{t}_{n})-\bm{g}(\bm{t}^{i}_{n}))\}\notag\\
=& 2\mathrm{Re}\{\bm{g}(\bm{t}^{i}_{n})^{H}\bm{C}_{n}\bm{g}(\bm{t}_{n})\}-\bm{g}(\bm{t}^{i}_{n})^{H}\bm{C}_{n}\bm{g}(\bm{t}^{i}_{n}).\notag
\end{align}
\end{subequations}
\indent  Then, we establish a positive real number denoted as $\gamma_{n}$, such that $\gamma_{n} \bm{I}_{2}\succeq \nabla^{2}\bar{z}(\bm{t}_{n})$. Similar to the derivation process in Problem (P2), we can obtain the following
\begin{small}
\begin{subequations}
\begin{align}
\bar{z}(\bm{t}_{n})=&\mathrm{Re}\{\bm{g}(\bm{t}^{i}_{n})^{H}\bm{C}_{n}\bm{g}(\bm{t}_{n})\}\tag{\theequation}\\
\geq & \bar{z}(\bm{t}^{i}_{n})+\nabla\bar{z}(\bm{t}^{i}_{n})^{T}(\bm{t}_{n}-\bm{t}^{i}_{n})-\frac{\gamma_{n}}{2}(\bm{t}_{n}-\bm{t}^{i}_{n})^{T}(\bm{t}_{n}-\bm{t}^{i}_{n})\notag\\
=& -\frac{\gamma_{n}}{2}\bm{t}_{n}^{T}\bm{t}_{n}+(\nabla\bar{z}(\bm{t}^{i}_{n})+\gamma_{n} \bm{t}^{i}_{n})^{T}\bm{t}_{n}-\nabla\bar{z}(\bm{t}^{i}_{n})^{T}\bm{t}^{i}_{n}\notag\\
&+\bar{z}(\bm{t}^{i}_{n})-\frac{\gamma_{n}}{2}(\bm{t}^{i}_{n})^{T}\bm{t}^{i}_{n}.\notag
\end{align}
\end{subequations}
\end{small}
\indent We can obtain:
\begin{subequations}
\begin{align}
&\mathrm{tr}(\bm{h}_{I}(\bm{t},\bm{r}_{0})\bm{h}_{I}(\bm{t},\bm{r}_{0})^{H} )\tag{\theequation}\\
=& -\gamma_{n}\bm{t}_{n}^{T}\bm{t}_{n}+2(\nabla\bar{z}(\bm{t}^{i}_{n})+\gamma_{n} \bm{t}^{i}_{n})^{T}\bm{t}_{n}-2(\nabla\bar{z}(\bm{t}^{i}_{n})^{T}\bm{t}^{i}_{n})\notag\\
&-\gamma_{n}(\bm{t}^{i}_{n})^{T}\bm{t}^{i}_{n}+\bm{g}(\bm{t}^{i}_{n})^{H}\bm{C}_{n}\bm{g}(\bm{t}^{i}_{n})+\bm{h}_{I}^{\bm{t}}(\bm{h}_{I}^{\bm{t}})^{H}.\notag
\end{align}
\end{subequations}
\indent Finally, in the $i$th iteration of SCA, we can obtain the following
\begin{subequations}\label{P5}
\begin{align}
\max_{\bm{t}_{n}}&~-\frac{\beta_{n}}{2}\bm{t}_{n}^{T}\bm{t}_{n}+(\nabla\bar{y}(\bm{t}_{n}^{i})+\beta_{n}\bm{t}_{n}^{i})^{T}\bm{t}_{n}
\tag{\theequation}\\
\text{s.t.}
&\quad \bm{t}_{n}\in C_{t},\label{P5:sub1}\\
&\frac{1}{\lVert \bm{t}_{n}^{i}-\bm{t}_{l} \rVert_{2}}(\bm{t}_{n}^{i}-\bm{t}_{l})^{T}(\bm{t}_{n}-\bm{t}_{l})\geq D,\label{P5:sub2}\\
&-\gamma_{n}\bm{t}_{n}^{T}\bm{t}_{n}+2(\nabla\bar{z}(\bm{t}^{i}_{n})+\gamma_{n} \bm{t}^{i}_{n})^{T}\bm{t}_{n}\notag\\
&-2(\nabla\bar{z}(\bm{t}^{i}_{n})^{T}\bm{t}^{i}_{n})-\gamma_{n}\bm{t}^{i}_{n})^{T}\bm{t}^{i}_{n}+\bm{g}(\bm{t}^{i}_{n})^{H}\bm{C}_{n}\bm{g}(\bm{t}^{i}_{n})\notag\\
&+\bm{h}_{I}^{\bm{t}}(\bm{h}_{I}^{\bm{t}})^{H}-\frac{\sigma_{I}^{2}\bar{\gamma}}{\mathrm{P}}\geq 0.\label{P5:sub3}
\end{align}
\end{subequations}
\indent The problem is formulated as a quadratically constrained quadratic program (QCQP). It can be acquired using the cplexqcp function\cite{b21}.
\begin{algorithm}
\caption{Alternating Optimization for Solving Problem (P1)}
\label{alg:A}
  \begin{algorithmic}[1]
    \STATE {Input: $\Sigma_{E}$, $\Sigma_{I}$, $\mathrm{P}$, $\sigma_{I}$, $M$, $N$, $L_{t}$, $L_{r}$, $\{\theta_{r}^{q}\}_{q=1}^{L_{r}}$,\\
    $\{\phi_{r}^{q}\}_{q=1}^{L_{r}}$, $\{\theta_{t}^{p}\}_{p=1}^{L_{t}}$, $\{\phi_{t}^{p}\}_{p=1}^{L_{t}}$, $C_{r}$, $C_{t}$, $D$, $\epsilon$.}
    \STATE {Initialize $\{\bm{r}\}$ and $\{\bm{t}_{n}\}_{n=1}^{N}$.}
    \STATE $\textbf{while}$ the increase of the power harvested in $\eqref{P1}$ is above $\epsilon$ $\textbf{do}$
    \STATE \hspace{0.2cm} Obtion the optimal solution for $\bm{Q}$ in Problem $(\mathrm{P1})$ given $\{\bm{t}_{n}\}_{n=1}^{N}$ and $\{\bm{r}\}$.
    \STATE \hspace{0.4cm} Obtain $\bm{A}_{n}$ via $\eqref{A}$.
    \STATE \hspace{0.4cm}Given $\bm{Q}$, $\{\bm{t}_{n}\}_{n=1}^{N}$, solve Problem $(\mathrm{P2})$ to update $\{\bm{r}\}$.
    \STATE \hspace{0.2cm} $\textbf{for}$ $n=1\rightarrow N$ $\bm{do}$
    \STATE \hspace{0.4cm}Obtain $\bm{B}_{n}$ via $\eqref{B}$.
    \STATE \hspace{0.4cm} Given $\bm{Q}$, $\{\bm{t}_{l}, l\neq N\}_{l=1}^{N}$, and $\{\bm{r}\}$ solve Problem $(\mathrm{P4})$ to update $\bm{t}_{n}$.
    \STATE \hspace{0.2cm} $\textbf{end for}$
    \STATE $\textbf{end while}$
    \STATE {Output $\bm{t}$, $\bm{r}$, $\bm{Q}$}.
  \end{algorithmic}
\end{algorithm}

\begin{figure}[h]
\centering
\includegraphics[width=6cm,height=4cm]{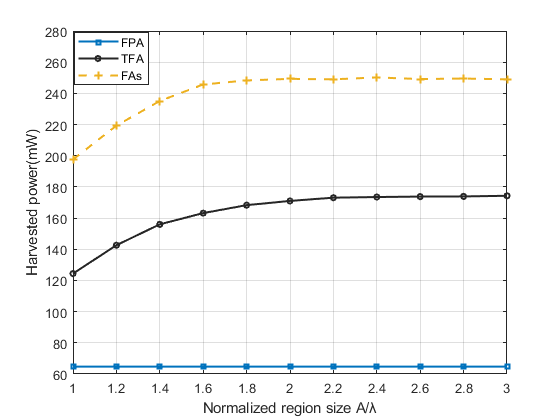}
\caption {Fluid antenna and fixed-position versus normalized region size.}
\label {2}
\end{figure}
\begin{figure}[h]
\centering
\includegraphics[width=6cm,height=4cm]{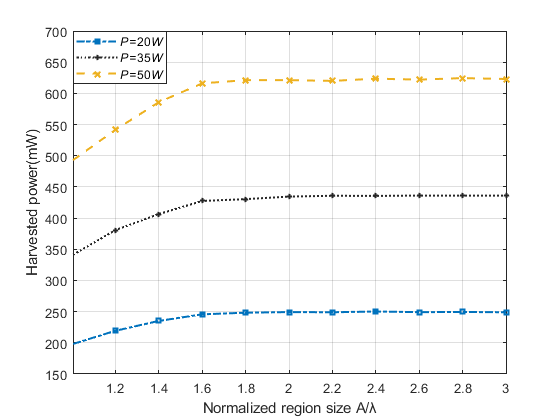}
\caption {Energy harvesting at different powers versus normalized region
size.}
\label {3}
\end{figure}
\begin{figure}[h]
\centering
\includegraphics[width=6cm,height=4cm]{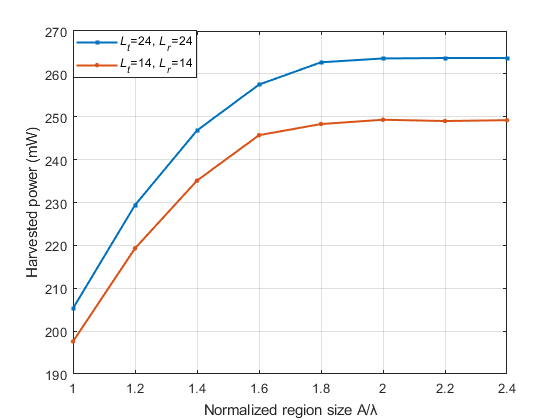}
\caption { FA systems with the number of transmit/receive paths $Lr = Lt = 14$, $Lr = Lt = 24$, versus normalized region
size.}
\label {4}
\end{figure}

\section{Numerical Results And Discussions}
In this section, numerical results is shown to validate the performance of the algorithm. In the simulation, we set $N$ = 4 transmit FAs , $M$ = 1 receive FA and $K$ = 1 receive fixed-position antenna. The base station transmit region and ER receive region are defined as square regions with dimensions $A\times A$. In this paper, we investigate transmission in a geometric channel model, where the count of transmit and receive paths is identical,
i.e.,$L_{t}=L_{r}$. Using $D=\lambda/2$ to represent the minimum distance constraint between fluidic antennas. The convergence threshold of the objective function is set to $\epsilon$.

We set $L_{t}=L_{r}=14$, $P$ = 20 W, $A = 3\lambda$, $\bar{\gamma}$ = 1 dB. As is shown in Fig.~\ref{2}, when FAs  are used in the receive and transmit regions, the EH efficiency is higher, while it is lowerer when FAs are used only in the transmit fluid antennas (TFA) and fixed-position antennas (FPA) are used in the receive region, scenarios above perform better compared with scenario where FPA are used in both the transmit and receive regions. This also demonstrates the superiority and better communication conditions of FAs.

We consider the performance tradeoff between base station transmission power and energy harvesting. In Fig.~\ref{3}, we assume that the SINR is 1 dB. As the power of the base station increases, the energy harvested by the ER shows a significant increase and tends to stabilize with the increase of the fluid antenna region.

We can infer from Fig.~\ref{4} that the ER can harvest more energy with the increase in the number of receive paths and transmit paths,
where the SINR is 1 dB. This also affirms how the communication rate of the fluid antenna is influenced by the number of paths.
\section{CONCLUSION}
In this paper, we introduce a novel FA-enabled MISO system designed for achieving simultaneous wireless information and power transfer, aiming to enhance EH by optimizing the antenna positions at the transmit/receive regions. We investigated the maximization of EH for FA-enabled point-to-point MISO communication, addressing the non-convex problem by decomposing it into three subproblems to ultimately find a locally optimal solution. The numerical results demonstrate a significant improvement in EH compared to traditional FPA-based MISO systems when using the proposed system and algorithm.

\end{document}